# Towards a finite conformal QED

A. D. Alhaidari

*Saudi Center for Theoretical Physics, P. O. Box 32741, Jeddah 21438, Saudi Arabia*

In 1986, while at UCLA working with C. Fronsdal and M. Flato, I proposed a model for conformal QED that I claimed to be divergence-free and nontrivial. The results for one loop calculation were given. However, a debate about unitarity and nontriviality of the model caused the withholding of publication of that work. Now and more than 30 years, students and colleagues suggested that with recent results there is a merit to the publication of the original study so that the problem could be revisited, independently investigated and the calculations be repeated. Consequently, I present here the work exactly as it appeared then in a UCLA preprint of the Theoretical Elementary Particle physics group with preprint number UCLA/86/TEP/31.

### TOWARD A FINITE CONFORMAL QED

A. D. Haidari
Department of Physics, University of California, Los Angeles, CA 90024

**ABSTRACT**: A finite (divergence-free) nontrivial conformal QED is formulated and the results of on-loop calculations are presented.

Recently [1], it was discovered that the conformal spinor field is a gauge field whose physical subspace is the massless electron. The representation of the conformal group carried by the spinor is non-decomposable and forms a symmetric triplet (the Gupta-Bleuler triplet). An indefinite-metric quantization, which is conformally covariant, was carried out in the same paper. Upon quantization of charged conformal fields, an associated difficulty arises [1-3]: No conformally covariant two-point function can be found that solves the free wave equation. The suggestion was made [3] that the "scalar" part of the triplet – the cyclic for the whole representation space – may resolve this problem when taken as an independent field component. In this paper, we study some of the implications of the findings made in Ref. [1] and write conformal QED in terms of the three components of the spinor gauge multiplet besides the usual vector potential and dipole ghost. The resulting theory eliminates the above-mentioned difficulty, but more dramatically, produces cancellation of divergent diagrams, verified here up to one loop.

We start by formulating the theory in Dirac's six-cone [4] notation, which has the advantage of manifest conformal invariance. The six-cone is Minkowski space compactified and embedded in $\mathbb{R}^6$ as the surface $y^2 = y_\alpha y_\beta \eta^{\alpha\beta} = y_0^2 - y_1^2 - y_2^2 - y_3^2 + 4y_+y_- = 0$ with the projection $\lambda y = y$, for $\lambda > 0$. The spinor action is [4]

$$S_0(\chi) = \int (dy)\left[-\tfrac{1}{2}\bar{\chi}\left(\overleftarrow{\not{\partial}}\not{y} + \not{y}\overrightarrow{\not{\partial}}\right)\chi + 2c\bar{\chi}\chi\right], \quad (dy) = \delta(y^2)d^6y \tag{1}$$

where $\chi$ is a homogeneous 8-component spinor with degree of homogeneity −2, and $c$ is a dimensionless real parameter. $\not{y} = y_\alpha \Gamma^\alpha$, where $\{\Gamma^\alpha\}$ is a set of 8×8 matrices satisfying the Clifford algebra $\{\Gamma^\alpha, \Gamma^\beta\} = 2\eta^{\alpha\beta}$ and as a basis we choose ($\{\gamma_\mu\}$ and $\vec{\sigma}$ are, respectively, the Dirac and Pauli matrices):

$$\Gamma^\mu = \sigma^3\gamma^\mu, \qquad \Gamma^\pm = \sigma^1 \pm i\sigma^2 = 2\sigma^\pm.$$

$\bar{\chi} = \chi^\dagger\Gamma$, where $\Gamma = \sigma^2\gamma^0$ and $\Gamma_\alpha^\dagger = -\Gamma\Gamma_\alpha\Gamma$. In Minkowski notation [1,4], $\chi$ forms the spinor multiplet $\hat{\psi}(x) \oplus \hat{\sigma}(x)$ using the transformation

$$\hat{\chi}(x) = (y^+)^2\left(1 - \sigma^+ x^\mu \gamma_\mu\right)\chi(y).$$

The most general homogeneous 2-point function is

$$\langle \chi(y)\bar{\chi}(y')\rangle = \left(y \cdot y'\right)^{-2} + c'\not{y}\not{y}'\left(y \cdot y'\right)^{-3}, \qquad c' \text{ is arbitrary.}$$

It carries the representation [1,5]:

$$\begin{bmatrix} D(5/2,1/2,0) \\ \oplus \\ D(5/2,0,1/2) \end{bmatrix} \to \begin{bmatrix} D(3/2,0,1/2) \\ \oplus \\ D(3/2,1/2,0) \end{bmatrix} \to \begin{bmatrix} D(5/2,1/2,0) \\ \oplus \\ D(5/2,0,1/2) \end{bmatrix}.$$

The arrows are semi-direct sums indicating leaks among the irreducible parts under the action of the group of conformal transformations. The left 2/3 of the triplet – the "scalar" and "physical" – is contained in $\hat{\psi}$, while $\hat{\sigma}$ is pure gauge and of the form $\not{y}\Omega$.

Aside from the trivial solution, no 2-point function solves the free wave equation obtained from (1) for all $c$ and $c'$. To try to resolve this problem, we split $\hat{\psi}$ into its "physical" and "scalar" parts in the following way:

$$\hat{\psi}(x) = \psi + \frac{c_1}{2}\not{\partial}\rho,$$
$$\hat{\sigma}(x) = \sigma + \frac{c_2}{2}\not{\partial}\psi + \frac{1+c_3}{4}\Box\rho,$$

where $c_i$ are dimensionless constants and $\not{\partial} = \gamma^\mu\partial_\mu$. To this end, we write the theory in terms of a homogenous spinor $\phi$ of degree −1, and make the identification:

$$\chi(y) \equiv \tfrac{1}{2}\bar{\partial}\phi(y) \qquad (\text{mod } y^2). \tag{2}$$

Now $\bar{\partial}$ is not intrinsic on the cone, therefore an extension off the cone [1,3,6] is needed for $\phi$ modulo $y^2 \not{y}$. In Minkowski notation, this extension takes the form of the multiplet $(\rho + y^2\sigma) \oplus \psi$ and definition (2) gives $c_1 = -c_2 = -c_3 = 1$. The action of special conformal transformation ($K_\mu$) on these spinors is

$$K_\mu \rho = \left( \overset{0}{\nabla}_\mu - \gamma_\mu \not{x} \right) \rho,$$

$$K_\mu \psi = \left( \overset{1}{\nabla}_\mu - \gamma_\mu \not{x} \right) \psi - \gamma_\mu \rho,$$

$$K_\mu \sigma = \left( \overset{2}{\nabla}_\mu - \gamma_\mu \not{x} \right) \sigma - \gamma_\mu \psi - \partial_\mu \rho,$$

where $\overset{n}{\nabla}_\mu = x^2 \partial_\mu - 2x_\mu \left( x \cdot \partial + n \right)$ and $\not{x} = x^\mu \gamma_\mu$.

The requirements of conformal covariance and the existence of a non-singular kernel (invertible propagator) give the following unique action and 2-point function:

$$S_0(\phi) = \int (dy) \Big[ \tfrac{1}{8} \big( \bar{\phi}\, \overleftarrow{\bar{\partial}} \big) \big( \overleftarrow{\bar{\partial}} \not{y} + \not{y} \overrightarrow{\bar{\partial}} \big) \big( \overrightarrow{\bar{\partial}} \phi \big) + \tfrac{1}{2} \big( \bar{\phi}\, \overleftarrow{\bar{\partial}} \big) \big( \overrightarrow{\bar{\partial}} \phi \big) - \tfrac{1}{8} \bar{\phi} \big( \overleftarrow{\partial}^2 \overleftarrow{\bar{\partial}} \not{y} + \not{y} \overrightarrow{\bar{\partial}} \overrightarrow{\partial}^2 \big) \phi \Big], \tag{3}$$

$$\mathcal{D}(y, y') = \left\langle \phi(y) \bar{\phi}(y') \right\rangle = \left( y \cdot y' \right)^{-1} + \tfrac{1}{2} \not{y}\not{y}' \left( y \cdot y' \right)^{-2} - \tfrac{1}{2} y^2 y'^2 \left( y \cdot y' \right)^{-3}, \tag{4}$$

satisfying the following equations (mod $y^2$):

$$\bar{\partial} \mathcal{D} = 0, \qquad \not{y} \partial^2 \mathcal{D} = 0. \tag{5}$$

Aside from the last term, whose origin will become clear shortly, this action is the same as (1) with $c = -1$, and in $x$-notation it reads:

$$-\int d^4x \Big[ \tfrac{\mathrm{i}}{2} \bar{\psi}\, \overleftrightarrow{\not{\partial}} \psi + \tfrac{\mathrm{i}}{2} \bar{\rho}\, \overleftrightarrow{\not{\partial}} \sigma + \tfrac{\mathrm{i}}{2} \bar{\sigma}\, \overleftrightarrow{\not{\partial}} \rho + 2\mathrm{i} \bar{\sigma} \psi - 2\mathrm{i} \bar{\psi} \sigma \Big]. \tag{6}$$

The non-vanishing covariant 2-point functions corresponding to (4) are:

$$\left\langle \rho(x) \bar{\rho}(x') \right\rangle = \frac{\mathrm{i}}{2\pi^2} \not{r} r^{-2}, \qquad \left\langle \rho(x) \bar{\psi}(x') \right\rangle = \frac{\mathrm{i}}{2\pi^2} r^{-2},$$

$$\left\langle \rho(x) \bar{\sigma}(x') \right\rangle = \frac{\mathrm{i}}{2\pi^2} \not{r} r^{-4}, \qquad \left\langle \psi(x) \bar{\psi}(x') \right\rangle = \frac{\mathrm{i}}{2\pi^2} \not{r} r^{-4}, \tag{7}$$

where $r_\mu = x_\mu - x'_\mu$ and $\not{r} = r^\mu \gamma_\mu$. These functions satisfy the free wave equations

$$\Box \partial\!\!\!/ \rho = 0, \quad \Box \psi = 0, \quad \partial\!\!\!/ \sigma = 0,$$

which also follow from (5) or (6).

Introducing minimal coupling ($\partial\!\!\!/ \to \partial\!\!\!/ + \mathrm{ie}a\!\!\!/$) in the original free action (1) gives

$$S(\chi) = \int (dy)\left\{ -\tfrac{1}{2}\bar{\chi}\left(\bar{\partial\!\!\!/} y\!\!\!/ + y\!\!\!/ \bar{\partial\!\!\!/}\right)\chi + 2c\bar{\chi}\chi - \tfrac{ie}{2}\bar{\chi}\left[y\!\!\!/, a\!\!\!/\right]\chi \right\}, \tag{8}$$

and $a_\alpha$ is the 6-vector potential which decomposes in $x$-notation into $(A_\mu, A_\pm)$ and has the following action [7]

$$S(a, j) = \int (dy)\left[\tfrac{1}{2}a^\alpha \partial^2 a_\alpha - \tfrac{\lambda}{8}(y\cdot a)\partial^4 (y \cdot a) - a \cdot j\right],$$

where $\lambda$ is a dimensionless gauge parameter. In the present model $y \cdot j = 0$, which can be used to eliminate $A_-$ and give the electromagnetic action [8]

$$S(A, J) = \int d^4x\left[-\tfrac{1}{4}F_{\mu\nu}F^{\mu\nu} + \tfrac{1-\lambda}{8}A_+\Box^2 A_+ - \tfrac{1}{2}A_+\Box\partial\cdot A - A_\mu J^\mu - 2A_+J_-\right].$$

The free propagators are [7,8]

$$\left\langle TA_\mu(x)A_\nu(x')\right\rangle = \frac{-1}{4\pi^2}r^{-2}\left[(\lambda-1)\eta_{\mu\nu} - 2\lambda r_\mu r_\nu r^{-2}\right],$$

$$\left\langle TA_\mu(x)A_+(x')\right\rangle = \frac{-1}{4\pi^2}r_\mu r^{-2}, \qquad \left\langle TA_+(x)A_+(x')\right\rangle = \frac{-1}{8\pi^2}. \tag{9}$$

The full spinor action (8) is invariant under the gauge transformation:

$$\chi \to e^{-\mathrm{ie}\Lambda}\chi; \qquad \text{alternatively, } \partial\!\!\!/ \phi \to e^{-\mathrm{ie}\Lambda}\partial\!\!\!/ \phi,$$

and in Minkowski notation it is equivalent to the following set of local gauge transformations

$$\partial\!\!\!/ \rho \to e^{-\mathrm{ie}\Lambda}\partial\!\!\!/ \rho, \quad \psi \to e^{-\mathrm{ie}\Lambda}\psi, \qquad \sigma \to e^{-\mathrm{ie}\Lambda}\left(\sigma - \tfrac{\mathrm{ie}}{2}\partial\!\!\!/ \Lambda\psi\right).$$

The unconventional affine gauge transformation of matter field is also present in conformal scalar QED [1]. In a gauge invariant perturbative quantum field theory defined by its n-point functions, one needs to define free propagators. If gauge transformation alters this definition, one fixes the gauge. Covariant gauge fixing is usually accomplished by adding a gauge-dependent term that vanishes on the physical subspace – i.e., the Lorentz condition term. For example, in QED we add $(\partial \cdot A)^2$ and in conformal QED $A_+\Box\partial\cdot A$ corresponding to the Lorentz condition $\partial \cdot A = 0$ and $A_+ = 0$ [7], respectively. For the present case, the Lorentz condition which removes the "scalar" mode from the spinor gauge field is $y\!\!\!/\phi = 0 = \rho$, and can only be imposed as an initial condition on the physical subspace. Therefore, we propose to fix the gauge in (8) by adding a term proportional to $\mathrm{i}\bar{\rho}\Box\overleftrightarrow{\not\partial}\rho = 0$ which leaves only rigid U(1) invariance. This exactly what has been done in a unique way under the requirements that lead to (3), actually the last term is nothing but $\frac{\mathrm{i}}{8}\bar{\rho}\Box\overleftrightarrow{\not\partial}\rho = 0$.

The interaction part of the Lagrangian density is

$$-\tfrac{\mathrm{i}e}{2}\bar{\chi}\left[\not\gamma,\not A\right]\chi = -\tfrac{\mathrm{i}e}{8}\left(\bar{\phi}\,\overleftrightarrow{\not\partial}\right)\left[\not\gamma,\not A\right]\left(\overleftrightarrow{\not\partial}\phi\right) = -\mathrm{i}eA^{\mu}\left(\bar{\psi}+\tfrac{1}{2}\bar{\rho}\,\overleftrightarrow{\not\partial}\right)\gamma_{\mu}\left(\psi+\tfrac{1}{2}\overleftrightarrow{\not\partial}\rho\right)$$
$$-\mathrm{i}eA_{+}\left(\bar{\psi}+\tfrac{1}{2}\bar{\rho}\,\overleftrightarrow{\not\partial}\right)\left(\sigma-\tfrac{1}{2}\overleftrightarrow{\not\partial}\psi\right)-\mathrm{i}eA_{+}\left(\bar{\sigma}-\tfrac{1}{2}\bar{\psi}\,\overleftrightarrow{\not\partial}\right)\left(\psi+\tfrac{1}{2}\overleftrightarrow{\not\partial}\rho\right)$$

The equation $\Box\not\partial\rho = 0$ is still maintained signifying that $\rho$ is a generalized free field and preserving the Lorentz condition. Arranging the three spinors in the triplet

$$\tau = \begin{pmatrix}\rho\\ \psi\\ \sigma\end{pmatrix} \quad \text{and} \quad \bar{\tau} = \begin{pmatrix}\bar{\rho} & \bar{\psi} & \bar{\sigma}\end{pmatrix},$$

the vertices are:

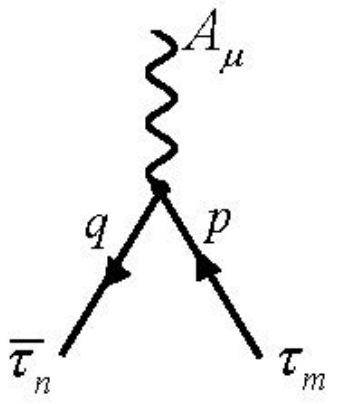

$$= e\begin{pmatrix} -\frac{\mathrm{i}}{4}\not q\gamma_{\mu}\not p & \frac{1}{2}\not q\gamma_{\mu} & 0\\ -\frac{1}{2}\gamma_{\mu}\not p & -\mathrm{i}\gamma_{\mu} & 0\\ 0 & 0 & 0\end{pmatrix}_{n,m}$$

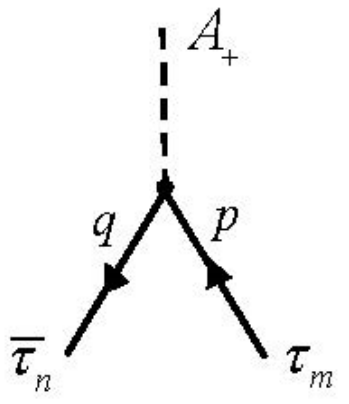

$$= e\begin{pmatrix} 0 & \frac{\mathrm{i}}{4}\not q\not p & \frac{1}{2}\not q\\ \frac{\mathrm{i}}{4}\not q\not p & \frac{1}{2}(\not p-\not q) & -\mathrm{i}\\ -\frac{1}{2}\not p & -\mathrm{i} & 0\end{pmatrix}_{n,m}$$

Using these vertices and the free propagators (7) and (9), one can demonstrate the finiteness of the theory by explicit calculations [9]. In this report, we state the results for one loop:

$\Sigma(p) =$ 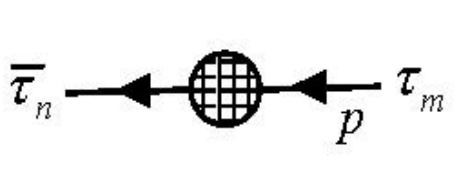

$$= \frac{e^2}{32\pi^2}\begin{pmatrix} \frac{\mathrm{i}}{4}p^2\not p & 0 & -\mathrm{i}\not p\\ 0 & -\mathrm{i}\not p & -2\\ -\mathrm{i}\not p & 2 & 0\end{pmatrix}_{n,m}$$

$\Lambda_{\mu}(p,q) =$ 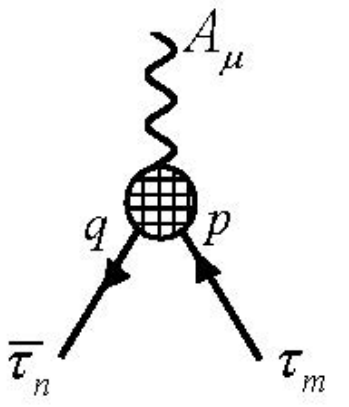

$$= \frac{-e^3}{32\pi^2}\begin{pmatrix} -\frac{\mathrm{i}}{4}\not q\gamma_{\mu}\not p & \frac{1}{2}\not q\gamma_{\mu} & 0\\ -\frac{1}{2}\gamma_{\mu}\not p & -\mathrm{i}\gamma_{\mu} & 0\\ 0 & 0 & 0\end{pmatrix}_{n,m}$$

$\Lambda_+(p,q) =$

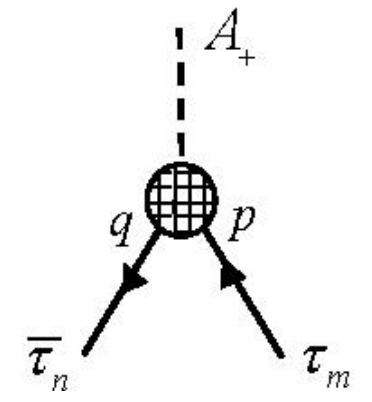

$$= \frac{e^3}{32\pi^2}\begin{pmatrix} \frac{1}{8}(q^2\not{p} - p^2\not{q}) & \frac{i}{4}q^2 & \frac{1}{2}\not{q} \\ \frac{i}{4}p^2 & 0 & -i \\ -\frac{1}{2}\not{p} & -i & 0 \end{pmatrix}_{n,m}$$

$\Pi_{\mu\nu}(r) =$

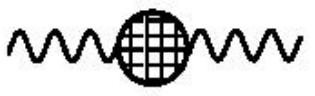

$$= 0$$

$\Pi_{\mu+}(r) =$

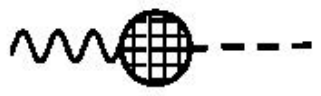

$$= 0$$

$\Pi_{++}(r) =$

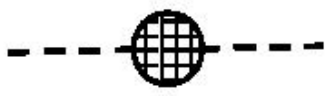

$$= 2e^2[\delta^4(r)]^2 = 2e^2\int(2\pi)^{-4}d^4q\,.$$

The last distribution is defined as follows: Let $F(r)$ and $G(r)$ be less singular distributions, then the insertion of $\Pi_{++}$ is given by

$$\int d^4x_2 d^4x_3 F(r_{12})\Pi_{++}(r_{23})G(r_{34}) \equiv 2e^2 F(r_{14})G(r_{14})\,,$$

which also leads to consistency with covariance. The Ward-Takahashi identity,

$$(p-q)^\mu \Lambda_\mu(p,q) = -e\left[\Sigma(p) - \Sigma(q)\right],$$

is satisfied on the physical subspace ( $\rho|\mathrm{phys}\rangle = 0$ ).

The first order contribution, $\Delta(p)$, to the free propagators is

$$\Delta_\tau(p) = \frac{-e^2}{32\pi^2}\begin{pmatrix} 0 & 0 & 0 \\ 0 & 0 & -1/2 \\ 0 & 1/2 & \frac{i}{4}\not{p} \end{pmatrix}$$

$$\Delta_{\mu\nu}(p) = \frac{-e^2}{4\pi^2}\left(-\eta_{\mu\nu} + 2p_\mu p_\nu p^{-2}\right)p^{-2}$$

$$\Delta_{\mu+}(p) = \frac{-e^2}{4\pi^2}\left(-2\,p_\mu / p^4\right)$$

$$\Delta_{++}(p) = \frac{-e^2}{4\pi^2}\left[-2\pi^2\delta^4(p)\right]$$

The author is indebted to C. Fronsdal for advice and helpful discussions and grateful to M. Flato for very stimulating and fruitful conversations. This work was supported in part by the Saudi Ministry of Higher Education and the University of Petroleum and Minerals.